**Accelerating CCSD(T) on Graphical Processing Units (GPUs)**

O. Jonathan Fajen,[1,2,3*] Joseph E. Kelly,[1*] Edward G. Hohenstein,[1,2,3] Todd J. Martinez[1,2,3]

[1]Department of Chemistry, Stanford University, Stanford, California 94305, United States

[2]The PULSE Institute, Stanford University, Stanford, California 94305, United States

[3]SLAC National Accelerator Laboratory, Menlo Park, California 94024, United States

## Abstract

Coupled cluster with singles, doubles and perturbative triples (CCSD(T)) often provides ground state correlation energies within 'chemical accuracy,' but suffers from high computational cost and steep scaling with system size. We present a GPU-accelerated implementation of CCSD(T) in the TeraChem software package. The new implementation achieves state-of-the-art performance, enabling the calculation of the (T) correction for a system with 63 atoms and more than 1000 basis functions in a little under 8 hours on a single node. Additionally, we demonstrate the utility of our optimized implementation for the rapid calculation of full CCSD(T)/CBS stacking energies for all ten unique DNA base pair stacked tetramers. We expect that the TeraChem CCSD(T) implementation will enable the rapid calculation of high-level data that was not previously accessible in a reasonable timeframe.

## Introduction

Coupled cluster (CC) theory has proven to be a powerful method for determining the ground state correlation energy of sufficiently single-reference systems.[1] As a wavefunction-based theory, CC offers a systematically improvable description of the system of interest, with the level of accuracy being essentially governed by the level of truncation in the cluster operator. The most popular flavor of CC is probably Coupled Cluster with Singles and Doubles (CCSD), which restricts the cluster operator to single and double excitations.[2] Higher order disconnected excitations are 'folded' into the CCSD wavefunction, due to the exponential ansatz. While CCSD often provides a qualitatively reasonable approximation to the wavefunction, at least some treatment of connected triples excitations is necessary to obtain thermochemical quantities with so-called 'chemical accuracy,' or errors of 1 kcal/mol errors or less. The simplest and most popular method for incorporating triples excitations into the ground state description is offered by Coupled Cluster Singles and Doubles with perturbative triples (CCSD(T)), often quoted as the 'gold-standard' in ground state quantum chemistry, which includes all the contributions to the 4$^{th}$ and 5$^{th}$ order correlation energy by the connected triples amplitudes projected onto the CCSD singles and doubles amplitude spaces.[3] For instance, CCSD(T) is often used to provide benchmark or reference values for higher order non-linear optical properties or non-covalent interactions, for small systems. [4-7]

Many of the weakest electronic interactions, such as the non-covalent interactions that govern the higher-order structures of proteins and DNA, are very sensitive to the inclusion of connected triples.[8] Unfortunately, it is not feasible to simply use CCSD(T) anytime the treatment of such small interactions is necessary, due to its daunting $O(N^7)$ scaling with system size. As a result, there have been many attempts to develop methods which obtain CCSD(T)-level accuracy at lower computational cost, mainly focusing on exploiting the local nature of electron correlation.[9-18] Pair natural orbital (PNO) methods stand out as perhaps the most successful of these, enabling the treatment of systems with hundreds of atoms with accuracy approaching that of CCSD(T). There has also been work on (T) corrections that leverage the rank-sparsity of the cluster amplitudes,[19-21] following similar approaches introduced for CC2 and CCSD.[22-34] These low rank approaches hold considerable promise, especially for electronically excited states where it is typically more difficult to find a local representation of electron correlation. There has also been a more recent surge of interest in methods that leverage a small amount of high-level data (eg, CCSD(T) energies) to train new force fields or create new density functionals that achieve near CCSD(T) level accuracy at only a fraction of the cost.[35-40] Despite the successes of lower-scaling approximate (T) methods and the rapid development of data-driven methods, the extension of the conventional CCSD(T) method to larger systems continues to be an important challenge, both in order to verify the accuracy of lower-level methods, and to provide training data for the smallest scale electronic interactions that are of increasing interest to practicing computational chemists. Reflecting this long-standing interest is a rich history of the development of optimized CCSD(T) codes, hand tailored for performance on some of the most powerful computers of their times. We briefly discuss several of the extant CCSD(T) implementations. More detailed reviews can be found in Refs.[41-43].

The CCSD(T) energy correction is calculated as contractions of two 6th-order tensors, $V_{ijk}^{abc}$ and $W_{ijk}^{abc}$, as in equation (1). The most expensive portion is the evaluation of $W_{ijk}^{abc}$, the scaling of which is dominated by an $o^3v^4$ term (where $o$ and $v$ are the numbers of occupied and virtual orbitals, respectively). Optimized parallelized implementations of the CCSD(T) algorithm distributed the evaluation of $W_{ijk}^{abc}$ along a subset of its indices. The first distributed implementation of a CCSD(T) calculation was described by Rendell, et al.[44] They only applied their parallelization strategies to the 4th order component of the (T) correction, but their reduction algorithms remain standard in modern fully distributed (T) codes. In particular, two of the algorithms they considered were denoted *abcijk* and *aijkbc*, in which reductions over all three virtual indices or over one occupied virtual pair (*ai*) are parallelized, respectively. The *abcijk* algorithm had several appealing properties, including a minimal per-node memory footprint ($o^3$), maximum task-parallelism, and near perfect scalability when replicating the full integral array. Unfortunately, replicating the integral array quickly became intractable and necessitated significant disk storage and an associated massive file I/O bottleneck. As a result, the *aijkbc* algorithm performed better, as its higher operation count (by a factor of four) was more than offset by a smaller I/O complexity.

Later, Kobayashi and Rendell implemented the first fully distributed version of CCSD(T) in NWChem, this time employing the *abcijk* algorithm.[45] Instead of storing the integrals and amplitudes on disk, they were kept in shared memory managed by the use of Global Arrays. This optimized implementation achieved a CCSD(T) calculation of a system with 1020 basis functions in ~2 hours on 90,000 processors.[46]

There have been several subsequent implementations that follow the *abcijk* distribution pattern, including in GAMESS,[47] PQS,[48] and MPQC.[42] The PQS CCSD(T) implementation achieved a calculation on a system with 1512 basis functions in ~72 hours on a 32-node commodity cluster. These algorithms tend to differ on the specifics of integral and amplitude replication but use the *abcijk* parallelization scheme due to its low node memory requirements and large number of parallel tasks. There are also implementations based on the *ijkabc* scheme, in GAMESS[49] and MRCC.[50, 51] The MRCC implementation was able to perform calculations on systems as large as 1569 basis functions on 224 cores. There is also a performant implementation in FHI-aims, although the distribution strategy employed there is not entirely clear.[52, 53] Finally, there are implementations which block over (distribute) all 6 indices, maximizing parallelism at the cost of increased communication overhead.[54, 55]

In the last 15 years or so, graphical processing units (GPUs) have become a standard hardware component of many high-performance quantum chemistry codes. The popularity of GPUs extends well beyond electronic structure theory, and more attention than ever is being paid to the development of algorithms that effectively leverage the vast computing power of GPU architectures. There has already been promising work in the GPU-acceleration of CCSD(T) algorithms, with the MPQC GPU (T) correction achieving a 3.6x speedup over the CPU implementation for a system with 60 atoms.[42] The NWChem TCE implementation also has GPU capability.[54] We also note impressive work on CCSD(T) implementations that scale over large, heterogenous architectures.[56, 57] In this work we describe a new GPU-based CCSD(T) algorithm implemented in a development version of TeraChem.[58-61] We show that this achieves state of the art performance for systems up to 1000 basis functions. Our implementation enabled the rapid study of DNA base pair stacking energies at the CCSD(T) level, a critical application for the development of ML-based force fields that was not previously accessible in a reasonable timeframe on a commodity cluster.

## Implementation Details

The (T) energy correction for a closed-shell singlet reference, assuming canonical orbitals, is:

$$E_{(T)} = \sum_{a\geq b\geq c} \left(2-\delta_{ab}-\delta_{bc}\right) \sum_{ijk} \left(W_{ijk}^{abc}+V_{ijk}^{abc}\right) \left(4W_{ijk}^{abc}+W_{jki}^{abc}+W_{kij}^{abc}-2W_{jik}^{abc}-2W_{ikj}^{abc}-2W_{kji}^{abc}\right) / D_{ijk}^{abc} \quad (1)$$

where

$$W_{ijk}^{abc} = P_{ijk}^{abc} \left( \sum_f G_{if}^{ab} t_{kj}^{cf} - \sum_l G_{ij}^{al} t_{lk}^{bc} \right) \quad (2)$$

$$V_{ijk}^{abc} = t_i^a G_{jk}^{bc} + t_j^b G_{ik}^{ac} + t_k^c G_{ij}^{ab} \quad (3)$$

$$D_{ijk}^{abc} = \varepsilon_a + \varepsilon_b + \varepsilon_c - \varepsilon_i - \varepsilon_j - \varepsilon_k \quad (4)$$

and P is the permutation operator such that

$$P_{ijk}^{abc} A_{ijk}^{abc} = A_{ijk}^{abc} + A_{ikj}^{acb} + A_{jik}^{bac} + A_{jki}^{bca} + A_{kji}^{cba} + A_{kij}^{cab} \quad (5)$$

$G_{pq}^{rs}$ is a two-electron integral in the molecular orbital basis, $t_i^a$ and $t_{ij}^{ab}$ are the converged CCSD amplitudes, $\varepsilon_i$ are the eigenvalues of the Fock matrix, *i*, *j*, *k*, and *l* index the *o* occupied orbitals and *a*, *b*, *c*, and *f* index the *v* virtual orbitals.

For all but the smallest systems/basis sets, the computational effort in CCSD(T) is dominated by the construction of the $W_{ijk}^{abc}$ tensor, which scales as $O(o^3v^4)$. Our GPU-accelerated implementation of the (T) correction follows the *abcijk* scheme of Rendell, et al.[44] The *abcijk* algorithm ensures a smaller memory footprint, at the cost of possibly requiring more blocks to be computed (looping through the three virtual indices). This is well-suited to the demands of GPU programming, where device memory may be quite small relative to the host (CPU), but device compute efficiency is extremely high. The sizes of tiles for virtual dimensions *a*, *b*, and *c* are given by $v_\alpha$, $v_\beta$, and $v_\gamma$, respectively. Tiles are sized to be as large as possible and still fit on the GPU. We require $v_\gamma \leq v_\beta \leq v_\alpha$, and for the largest systems possible, the length of this virtual index can shrink to one. Elements within a tile are indexed by compound index $\alpha\beta\gamma ijk$, where $\alpha \leq v_\alpha$, $i \leq o$, and so on. This best-case scenario memory footprint therefore scales as $O(o^3)$. Conventional high-performance CCSD implementations (including the one used in TeraChem[62]) typically store the full $o^2v^2$ doubles amplitudes in CPU memory. When the doubles amplitudes become larger than ~200 GB (for instance, with $o = 150, v = 1150$, the doubles amplitudes are already 238 GB), the doubles amplitudes must be stored in part on disk, significantly degrading performance for those CCSD implementations which do not simply fail for systems of this size. The ideal $O(o^3)$ memory consumption of the CCSD(T) implementation is not even 1 GB for such a system. From a memory footprint perspective, the systems for which we can now perform CCSD calculations are therefore easily accessible by CCSD(T).

In GPU-programming, it is important to pre-allocate memory on the device and reuse the same space, rather than dynamically allocating and deallocating memory, which can lead to fragmentation and apparent lack of memory. Therefore we use pre-allocation rather than dynamic memory management, avoiding slowdowns stemming from fragmentation while enabling asynchronous memory transfer and improving latency. In line with the CCSD method in TeraChem, the electron repulsion integrals (ERIs) are approximated by a Cholesky Decomposition (CD)[63-65]

$$G_{pq}^{rs} \approx \sum_X L_{pq}^X L_{rs}^X \tag{6}$$

where $X$ runs over the auxiliary Cholesky basis. Pieces of the Cholesky vectors are moved to GPUs and the ERIs are then reconstructed on the GPU as needed. This helps reduce memory transfer time, a major bottleneck when using GPUs. All arrays with more than one virtual index are tiled, meaning that only one index is of length $v$, while other virtual dimensions are of length $v_\alpha$. Therefore, the only arrays which are replicated in full across all GPUs are the orbital energies (of length $o + v$), the singles amplitudes, $t_i^a$, and the occupied-occupied Cholesky array, $L_{ij}^X$. The steepest scaling terms that must be stored on each GPU are the components of $W_{ijk}^{abc}$ and $V_{ijk}^{abc}$, which are of size $v_\alpha^3 o^3$. Figure 1 outlines the tiling strategy used for handling the $W_{ijk}^{abc}$ tensor. Each GPU receives a batch of tiles to compute, where each tile is a submatrix, $W_{ijk}^{\alpha\beta\gamma}$, of the full $W_{ijk}^{abc}$. The bulk of resources are dedicated to building and storing the W tensors. Custom built CUDA kernels make permuting those arrays efficient and are essential for minimizing the number of copies stored simultaneously and improving the efficiency of access to those arrays when accumulating the energy contributions.

To achieve the fastest results, we moved as much of the computation into matrix multiplies as possible and used the highly optimized NVIDIA cuBLAS library. We permute several Cholesky vectors on the CPU because the CCSD and (T) components benefit from different index orderings. Pseudocode for the full CCSD(T) energy calculation is shown in Figure 2. Each GPU receives a batch of virtual indices to compute, of lengths $v_\gamma \le v_\beta \le v_\alpha$. The first step is to copy over the necessary parts of the occupied-virtual and virtual-virtual Cholesky vectors (lines 2 and 3 in Figure 2). Next, we loop through the six permutations needed to build $W_{ijk}^{\alpha\beta\gamma}$ (Equation 2). We form the $W_{ijk}^{\alpha\beta\gamma}$ tensor in two stages, corresponding to the two terms in Equation 2. Two copies of the doubles amplitudes are stored on the device (GPU), one with both virtual indices tiled, and another with only one virtual index tiled. The first contraction, $G_{if}^{\alpha\beta} t_{kj}^{\gamma f}$, is performed as a DGEMM between arrays of shapes $(v_\alpha v_\beta o, v)$ and $(v, v_\gamma o^2)$, which leads to a contribution to $W_{ijk}^{\alpha\beta\gamma}$ that we denote as $W_v$ and has an indexing of $\alpha\beta i\gamma jk$ after performing the contraction. The second contraction, $G_{ij}^{\alpha l} t_{lk}^{\beta\gamma}$, is computed as a DGEMM between shapes $(v_\alpha o^2, o)$ and $(o, v_\beta v_\gamma o)$. This leads to a contribution denoted $W_o$ and is initially indexed as $\alpha ij\beta\gamma k$. We form these two contributions separately and accumulate them into a final array indexed as $\alpha\beta\gamma ijk$, meaning that we store two copies of the $v_\gamma v_\beta v_\alpha o^3$-sized $W_{ijk}^{\alpha\beta\gamma}$ arrays. The accumulation involves permuting the indices of the temporary array ($W_o$/$W_v$), which is accomplished using a custom CUDA kernel, pseudocode for which can be found in Figure 3.

An important detail of our implementation is the use of *cudaMemcpyAsync()* calls in the construction of the $W_{ijk}^{\alpha\beta\gamma}$ components. The construction of these components requires building a set of two-electron integrals (lines 7 and 11 in Figure 2) and then contracting the integrals with a segment of the doubles amplitudes (lines 8 and 12 in Figure 2). By using *cudaMemcpyAsync()* calls to copy the next needed piece of the doubles amplitudes, we can perform the Cholesky factor contraction while the copy occurs, meaning that when we have formed, for instance, $G_{ij}^{\alpha l}$, the needed doubles amplitudes, $t_{lk}^{\beta\gamma}$, are already available on the GPU and we can proceed to the next contraction.

Once we have constructed the permutations of $W_{ijk}^{\alpha\beta\gamma}$, we construct the two-electron integrals needed to build $V_{ijk}^{\alpha\beta\gamma}$ (Equation 6 and lines 16-18 in Figure 2). The Cholesky vectors we need for this step are already on the GPU (copied over in lines 2-3), so there is no further CPU/GPU memory transfer. With these integrals in hand, we loop through the pieces of the virtual indices assigned to the GPU and compute their contributions to the (T) energy correction (lines 19-25 in Figure 2). Pseudocode for the energy kernel is shown in Figure 4. Each thread is assigned a set of occupied indices according to its place in the block (see lines 4-6 in Figure 4). Next, we set up the indices that we will use later. Then, we build the orbital energy denominator according to Equation 4 in line 23 of Figure 4 and we build $V_{ijk}^{\alpha\beta\gamma}$ according to Equation 3 in line 24 of Figure 4. After accumulating the permutations of $W_{ijk}^{\alpha\beta\gamma}$ in lines 25-30, we compute the contribution to the energy of the index set $\alpha\beta\gamma ijk$ according to Equation 1 (line 31 of Figure 4).

**Results**

We begin by demonstrating the strong scaling and parallel efficiency of our GPU-accelerated (T) correction on a set of water clusters of increasing size with the cc-pVDZ basis set. As shown in Figure 5, we are able to achieve nearly ideal strong scaling across 1 to 8 GPUs, even for the smallest water clusters studied. Of course, as the system size increases, more and more of the computational effort is spent on the GPUs and we observe better scaling. The strong scaling results for small system sizes demonstrate near ideal load-balancing in our implementation, and this is also reflected in the parallel efficiency we observe. Figure 6 shows that we observe a parallel efficiency greater than 0.9 on 8 cards, even for the smallest water cluster (15 atoms). For the largest clusters (105 atoms), the parallel efficiency is essentially 1.0 for 2, 4, or 8 GPUs.

For systems of small to medium size, such as the water clusters treated here, it is very often the case that an optimized (T) correction can actually be cheaper than the underlying CCSD calculation, particularly since the (T) correction is non-iterative and the CCSD equations are eventually limited by cheap but poorly parallelizable terms, as shown by Datta and Gordon.[49] This is the case for our GPU-accelerated CCSD(T) implementation, as seen in Figure 7. The (T) correction only becomes the majority of the compute time for the largest system considered, 35 water molecules. Eventually, the higher scaling of the (T) term overpowers the poor computational efficiency of the CCSD calculation.

Our implementation also scales very well with basis set size. We performed calculations on 14 waters using 1 and 8 A100s with several basis sets: cc-pVDZ, aug-cc-pVDZ, cc-pVTZ, and aug-cc-pVTZ.[66, 67] This led to basis sets ranging from 350 to 1470 functions. In addition to the absolute size of calculation increasing, this tests how the code behaves as the ratio of virtual to occupied orbitals increases. The number of tiles in the calculation increases significantly since we tile over the three virtual indices and the size of the Cholesky vectors grows as well. In all cases, the parallel efficiency from 1 to 8 GPUs was above 85%. Figure 8 shows that the (T) calculation has yet to grow so large that it dominates the computational time. In the case of the largest aug-cc-pVTZ calculation with 1470 basis functions, the (T) correction is only 37% of the total CCSD(T) calculation time.

Although the majority of optimized CCSD(T) codes use CPUs, we can compare our (T) implementation in TeraChem with the GPU-based implementation in MPQC.[42] In Table 1 we compare the (T) time for $(H_2O)_{14}$/cc-pVDZ with TeraChem and MPQC on 1 or 2 NVIDIA P100 GPUs. With 2 P100s, the MPQC (T) time is 39 minutes. In contrast, the same calculation in our implementation using 2 P100s takes 18.5 minutes.

We compare our GPU-accelerated timings against the CCSD(T) implementations in packages including MPQC, MRCC, ORCA, Psi4, GAMESS, and FHI-aims which use multi-threaded, shared memory, CPU-based implementations. It is important to note that the comparisons are taken from previous reports in the literature, and therefore a wide variety of compilation environments and hardware are represented. This makes direct comparisons of performance difficult. These statistics are provided to give the reader a qualitative sense of the state of the art for CCSD(T) computations and not for a quantitative comparison. Interested readers are referred to Table S1 in the SI for details about the various CPU/GPU hardware used for the data extracted from literature reports.

A relevant performance metric for dataset generation is node-hours, indicating the cost in terms of both time and compute resources consumed, which we show in the last column of Table 2. The results in Table 2 show that our GPU implementation makes CCSD(T) calculations more routine, with roughly an order of magnitude decrease in the number of node-hours required for these three example systems. As the calculations get bigger, they take more complete advantage of A100 compute resources (Figure S1) and as a result the performance edge of our GPU-implementation grows. Calculating the CCSD(T) ground state energy for a single configuration of any of the molecular systems shown here is easily accessible for any of these codes. However, a key use case for high accuracy coupled cluster calculations is the generation of large datasets of reference data for machine learning applications.[38-40] The ability to generate roughly 10 times or more training data using a GPU-accelerated method compared to a purely CPU-based implementation could make training robust machine learning algorithms significantly easier.

As shown in Table 3, the speed-ups achievable through GPU-acceleration are even more dramatic as the molecular size increases. Again, the number of node-hours required with our GPU implementation is consistently at least an order of magnitude less than a variety of other CPU-based codes. In one of the largest tests, we compared MPQC, NWChem and TeraChem using a 2-nucleotide strand of DNA neutralized with a sodium ion: GC-dDMP-B. The NWChem calculation used 20,000 nodes, each with 2 cores and finished in 84 minutes. While the total wall time was very short, the calculation consumed 28,000 node-hours. MPQC was able to achieve a 9x improvement in terms of node-hours, with 64 nodes performing the calculation in just over 3000 node-hours. However, the total wall time was more than two days for the (T) calculation. With 8 A100s in one node, our implementation finishes this calculation in under 8 hours.

**Applications**

Our GPU-accelerated CCSD(T) enables analyses that were previously difficult or impractical on a commodity cluster. As an example, we examine the set of all possible DNA nucleotide tetramers arranged with two base-pairs stacked on top of each other. Structures are taken from the energy minima described by Parker et. al. (see Tables 42-50 in the SI of their work).[68] All geometries, including water clusters and nucleotide monomers, dimers, trimers, and tetramers, can be found in the SI of this paper. The labelling of these structures is as follows. X•Y means an X base is hydrogen-bonded to a Y base. WX:YZ means X•Y is stacked on top of W•Z akin to their orientation in a DNA strand. With the 4 base pairs of DNA, there are 16 possible structures. However, due to symmetry, only 10 of them are unique.

The efficiency of our implementation enables complete CCSD(T)/aug-cc-pVDZ energy calculations on all possible monomers, dimers, trimers, and the full tetramer for each of the 10 structures. We note that each calculation in this dataset is comparable in size to the GC-dDMP-B calculation, meaning that evaluating this dataset at the CCSD(T)/aug-cc-pVDZ level requires performing some of the largest known CCSD(T) calculation tens of times. With our GPU-accelerated CCSD(T) implementation in TeraChem, this task is eminently feasible. For example, the full CCSD(T)/aug-cc-pVDZ calculation of GG:CC (58 atoms, 1130 basis functions) on one node with 8 A100 GPUs takes about 14.2 hours and spends about 6.6 hours to carry out the (T) component. The (T) calculation runs at 90% peak efficiency of the general double precision performance, comparable to the performance from the water cluster tests (see Figure S1). We used a previously described method[69] to estimate the complete basis set limit (CBS) correction, in which the MP2 correlation energy is extrapolated to the CBS limit using the method of Halkier, et al.,[70] and the CCSD(T) CBS extrapolation is given by correcting the CBS MP2 correlation energy by the difference between MP2 and CCSD(T) in a smaller basis set (aug-cc-pVDZ). MP2 calculations were performed with Psi4.[71] In short:

$$\Delta E_{CBS}^{CCSD(T)} \approx \Delta E_{CBS}^{MP2} + \left(\Delta E_{X}^{CCSD(T)} - \Delta E_{X}^{MP2}\right) \tag{7}$$

where the subscript denotes the basis set being used and the cardinal index $X$ is set to two. The MP2 correlation energy is extrapolated to the CBS limit as

$$E_{CBS}^{MP2} \approx E_{X}^{MP2} - BX^{-3} \tag{8}$$

where the cardinal index $X$ is set to four. To avoid basis set superposition error (BSSE), we carry out all calculations from the monomers to the tetramer with the same set of basis functions necessary for the complete tetramer. Previous studies have examined the stabilization energy due to stacking two base pairs but were limited to using methods like ΔCCSD(T) to estimate the higher order interactions (only the dimers were calculated with CCSD(T), while trimers and tetramers were calculated with MP2).[69, 72] The stacking energies are defined as

$$\Delta E^{WX:YZ} = E^{WX:YZ} - \left(E^{W\bullet Z} + E^{X\bullet Y}\right) - \left(E^{W} + E^{X} + E^{Y} + E^{Z}\right) \tag{9}$$

Table 4 lists these previous approximate results for these stacking energies along with our CCSD(T)/CBS results from the complete set of CCSD(T)/aug-cc-pVDZ calculations for all tetramers and MP2/aug-cc-pV5Z calculations to estimate the CCSD(T)/CBS results using Eqs. 7 and 8. In general, the extrapolation methods tend to underestimate the stabilization energy due to stacking compared to the full CCSD(T)/CBS calculations, although the magnitudes of the deviations are very small. In general, the level of agreement reflects well on the ability of approximate methods to reproduce near-benchmark level accuracy.

The tendency for approximate methods to underestimate the stabilization from stacking can also be seen in our calculations of the non-additive component of the 4-body interaction energy at different levels of theory in the lower panel of Figure 9. The non-additive 4-body component of the interaction energy is calculated as

$$\begin{aligned}\Delta E^{4-body} = E^{WX:YZ} &- \left(E^{WXZ} + E^{XYZ} + E^{WYZ} + E^{WXY}\right) - \left(E^{W} + E^{X} + E^{Y} + E^{Z}\right) \\ &+ \left(E^{W\bullet Z} + E^{W\bullet Y} + E^{W\bullet Z} + E^{X\bullet Y} + E^{X\bullet Z} + E^{Y\bullet Z}\right)\end{aligned} \tag{10}$$

One can similarly define a non-additive 3-body contribution from the WXY trimer to the WX:YZ stacking energy as

$$\Delta E^{3-body} = E^{WXY} - \left(E^{W\bullet X} + E^{W\bullet Y} + E^{X\bullet Y}\right) \tag{11}$$

The upper panel of Figure 9 shows the average non-additive 3-body contribution, obtained by averaging over all four constituent trimers of WX:YZ. Progressing from MP2 to CCSD to CCSD(T), the 4-body interaction becomes more stabilizing. The MP2/CBS 4-body contribution does not always even point in the same direction as CCSD(T)/CBS and often substantially underestimates the CCSD(T)/CBS value, while the CCSD/CBS 4-body values are generally quite close to those at the CCSD(T)/CBS level. The 3-body interactions are always destabilizing and, on average, become more so with higher levels of theory. These are potentially contributions from dispersion effects which CCSD(T) is known to capture well, but which pose a challenge for MP2 and CCSD.

**Conclusions**

We have presented a GPU-accelerated implementation of CCSD(T) which achieves state-of-the-art performance for calculations on systems up to more than 1000 basis functions. Even for the largest system studied, the (T) correction is faster than the underlying CCSD calculation, due in part to the excellent parallel efficiency and peak performance we achieve in our implementation. Further, we are able to rapidly perform an analysis of interaction energies between DNA monomers at an unprecedented level of accuracy. This should enable rapid generation of benchmark-level data for potential machine-learning methods. Furthermore, our implementation can be deployed on a commodity cluster and still achieve better performance than massively parallel implementations on institutional clusters. Whereas a recent set of CCSD(T) calculations of roughly 8000 non-equilibrium noncovalent interactions was generated in about 1.5 million node hours,[37] we project a total cost of ~70,000 node hours (> 20x less) with our implementation.[73] Another avenue for increasing the regime of applicability of CCSD(T)-level calculations will be the synthesis of our GPU-implementation with tensor hypercontracted and/or rank-reduced approaches, which can lower the formal scaling of the method. Indeed, a Rank-Reduced formulation of CCSD(T) has already been introduced and demonstrated excellent performance for systems up to at least 50 atoms.[20]

**Supporting Information**

Hardware specifics for timings of CCSD(T) implementations presented in the text, as well as operation rate as measured in TFLOPs for our CCSD(T) implementation across system size and on both 1 and 8 A100 and V100 GPUs.

**Data Availability**

Inputs and outputs for all of the TeraChem and Psi4 calculations can be found at: https://dx.doi.org/10.5281/zenodo.17756230.

**Acknowledgements**

This work was supported by the AMOS program of the U.S. Department of Energy, Office of Science, Basic Energy Sciences, Chemical Sciences, Geosciences, and Biosciences Division. OJF is a Department of Energy Computational Science Graduate Fellow (Award Number DE-SC0023112), supported by the U.S. Department of Energy, Office of Science, Office of Advanced Scientific Computing Research.

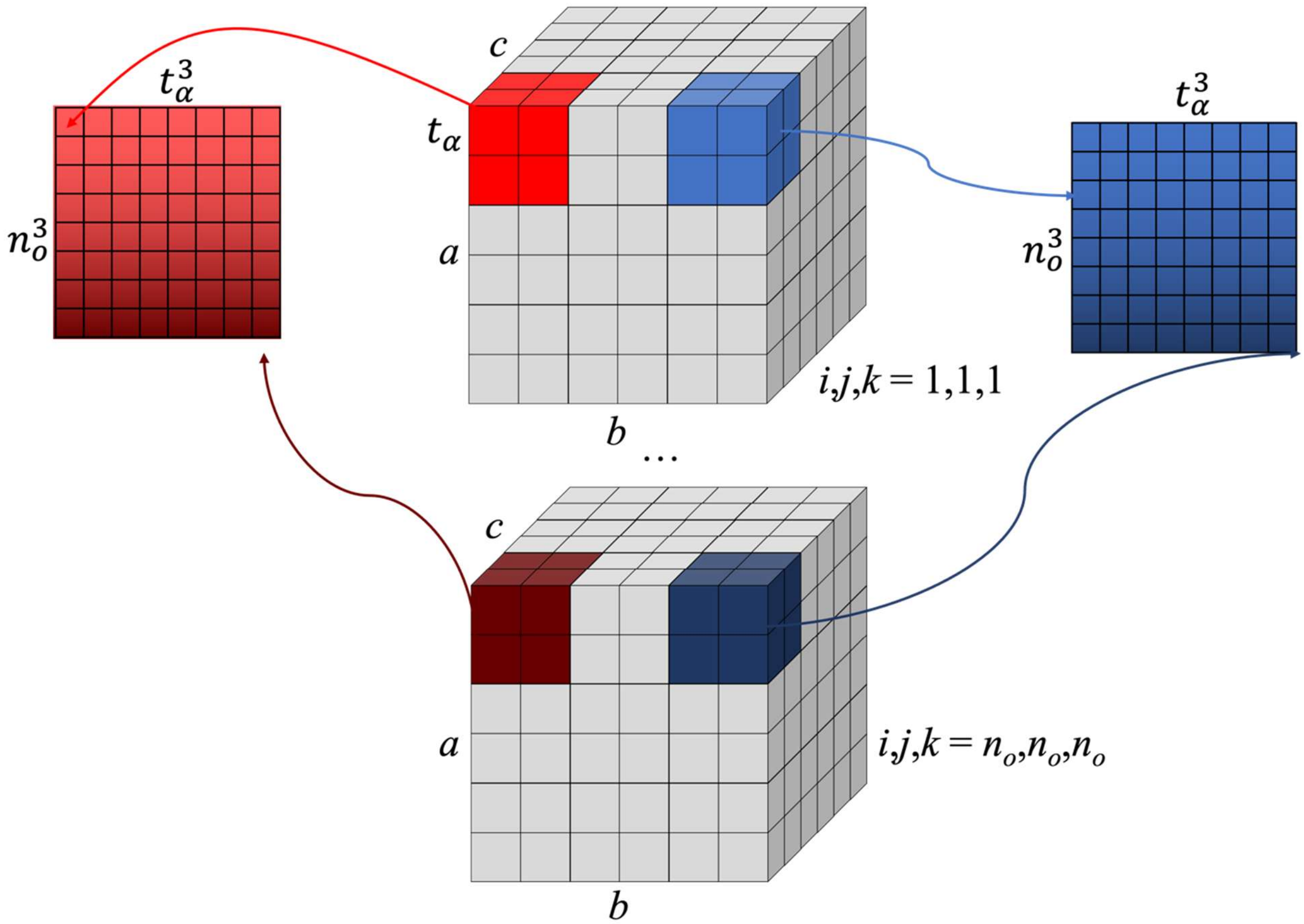

**Figure 1**. Schematic outlining the tiling procedure for $W_{ijk}^{abc}$. A total of $v_\alpha v_\beta v_\gamma$ strips of length $o^3$ are computed in each batch. The full $W_{ijk}^{abc}$ is represented by $o^3$ cubes, each $v$ by $v$ by $v$. The color gradient from light to dark represents the progression of the compound occupied dimension. The submatrix that is distributed to one GPU, $W_{ijk}^{\alpha\beta\gamma}$, where $\alpha \leq v_\alpha$ ,$\beta \leq v_\beta$, $\gamma \leq v_\gamma$ is constructed from a sub-cube of size $v_\alpha v_\beta v_\gamma$ from each of the $o^3$ cubes comprising $W_{ijk}^{abc}$. The contribution of each submatrix, $W_{ijk}^{\alpha\beta\gamma}$, to the (T) energy is calculated separately.

**GPU-acclerated CCSD(T) Core Algorithm**

1: Input: $v_\alpha \geq v_\beta \geq v_\gamma$
2: Copy from $L^X_{ai}$ to $L^X_{\nu i}$ for $\nu$ in ($\alpha$, $\beta$, $\gamma$) (size $v_\nu oX$)
3: Copy from $L^X_{ab}$ to $L^X_{\nu b}$ for $\nu$ in ($\alpha$,$\beta$,$\gamma$) (size $v_\nu vX$)
4: ▷ Loop through permutations in Equation 5
5: **for** ($\alpha$,$\beta$,$\gamma$,$i$,$j$,$k$) in $P^{\alpha\beta\gamma}_{ijk}$ **do**
6: Copy from $t^{ab}_{ij}$ to $t^{\beta\gamma}_{lk}$ (size $o^2 v_\beta v_\gamma$)
7: Build $G^{\alpha i}_{jl} = L^X_{jl}\ L^X_{\alpha i}$ ▷ Equation 6
8: Build $(W_o)_{\alpha ij\beta\gamma k} = G^{\alpha i}_{jl}\ t^{\beta\gamma}_{lk}$ (DGEMM) ▷ Second term in Equation 2
9: $W_{\alpha\beta\gamma ijk}$ = **PERMUTE**($(W_o)_{\alpha ij\beta\gamma k}$)
10: Copy from $t^{ab}_{ij}$ to $t^{\gamma f}_{jk}$ (size $v_\gamma vo^2$)
11: Build $G^{\alpha i}_{\beta f} = L^X_{\beta f}\ L^X_{\alpha i}$ ▷ Equation 6
12: Build $(W_v)_{\alpha i\beta\gamma jk} = G^{\alpha i}_{\beta f}\ t^{\gamma f}_{jk}$ (DGEMM) ▷ First term in Equation 2
13: $W_{\alpha\beta\gamma ijk}$ += **PERMUTE**($(W_v)_{\alpha i\beta\gamma jk}$)
14: **end for**
15: ▷ Construct integral arrays needed to build $V^{\alpha\beta\gamma}_{ijk}$ ▷ Equation 3
16: Build $G^{\alpha\beta}_{ij} = L^X_{\alpha i}\ L^X_{\beta j}$ ▷ Equation 6
17: Build $G^{\beta\gamma}_{jk} = L^X_{\beta j}\ L^X_{\gamma k}$ ▷ Equation 6
18: Build $G^{\alpha\gamma}_{ik} = L^X_{\alpha i}\ L^X_{\gamma k}$ ▷ Equation 6
19: **for** $\gamma \leq v_\gamma$ **do**
20: **for** $\beta \leq v_\beta$ **do**
21: **for** $\alpha \leq v_\alpha$ **do**
22: Compute **triples_energy**($G^{\alpha\beta}_{ij}$,$G^{\beta\gamma}_{jk}$,$G^{\alpha\gamma}_{ik}$,$P^{\alpha\beta\gamma}_{ijk} W^{\alpha\beta\gamma}_{ijk}$)
23: **end for**
24: **end for**
25: **end for**

**Figure 2.** Pseudocode for the core CCSD(T) energy computation. Each GPU receives a batch of virtual indices with lengths $v_\gamma \leq v_\beta \leq v_\alpha$. Arrays stored on device (GPU) memory are written in green text, while arrays stored in global (CPU) memory are written in red text. The first step is to copy Cholesky vectors corresponding to each of the tiles. Next, we construct the portions of $W^{abc}_{ijk}$, indexed as $W^{\alpha\beta\gamma}_{ijk}$, by looping through the permutations in Equation 5. The Copy calls within the permutation loop are performed with cudaMemcpyAsync() calls, which allow us to copy over the next needed piece of the doubles amplitudes while the next integral array is constructed (see, eg, lines 8 and 9). As discussed in the text, the two terms of Equation 2 are formed separately ($W_o$ and $W_v$ in lines 10 and 14), and then permuted into the final accumulated array (lines 11 and 15). Pseudocode for the permutation can be found in Figure 3. After the contributions to $W^{\alpha\beta\gamma}_{ijk}$ have been formed, we construct the integral arrays needed to build $V^{\alpha\beta\gamma}_{ijk}$ in Equation 3 (lines 18-20). Finally, we compute the contribution of each $(\alpha, \beta, \gamma)$ triplet in line 24. The energy kernel pseudocode is shown in Figure 4.

**PERMUTE**

1: Input: $W_{\alpha i\beta\gamma jk}$, $blockIdx$, $blockDim$, $threadIdx$
2: Output: $W_{\alpha\beta\gamma ijk}$
3: $\alpha ij = blockIdx.z * blockDim.z + threadIdx.z$
4: $\beta\gamma = blockIdx.y * blockDim.y + threadIdx.y$
5: $k = blockIdx.x * blockDim.x + threadIdx.x$
6: $\alpha i = \alpha ij/o$
7: $j = \alpha\%o$
8: $\alpha = \alpha i/o$
9: $i = \alpha i\%o$
10: $\beta = \beta\gamma/v_\beta$
11: $\gamma = \beta\gamma\%v_\beta$
12: $\alpha i\beta\gamma jk = (\alpha ij * v_\beta * v_\gamma + \beta * \gamma) * o + k$
13: $\alpha\beta\gamma ijk = ((\alpha * v_\beta * v_\gamma + \beta * \gamma) * o^2 + ij) * o + k$
14: **if** $\alpha ij < v_\alpha * o^2$ && $\beta\gamma < v_\beta * v_\gamma$ && $k < o$ **then**
15: $\quad W_{\alpha\beta\gamma ijk}[\alpha\beta\gamma ijk] = W_{\alpha i\beta\gamma jk}[\alpha i\beta\gamma jk]$
16: **end if**
17: Output: $W_{\alpha\beta\gamma ijk}$

**Figure 3.** Pseudocode of the permutation kernel. Each thread is assigned an index set (values of $\alpha\beta\gamma ijk$) according to its position in the block (lines 3-11). Next, the old and new compound indices are calculated (lines 12 and 13). Finally, the permutation is performed in line 14. As before, arrays stored in device (GPU) memory are shown in green text.

**triples_energy**

1: ▷ $\epsilon_a$, $\epsilon_i$, $t_i^a$ are replicated across all GPUs

2: Input: $\alpha$,$\beta$,$\gamma$,$G_{ij}^{\alpha\beta}$,$G_{jk}^{\beta\gamma}$,$G_{ik}^{\alpha\gamma}$, $P_{ijk}^{\alpha\beta\gamma} W_{ijk}^{\alpha\beta\gamma}$, $\epsilon_a$, $\epsilon_i$, $t_i^a$, $blockIdx$, $blockDim$, $threadIdx$

3: Output: $sum_{ijk}$

4: $i = blockIdx.x * blockDim.x + threadIdx.x$

5: $j = blockIdx.y * blockDim.y + threadIdx.y$

6: $k = blockIdx.z * blockDim.z + threadIdx.z$

7: ▷ Set up virtual indices

8: $\alpha\beta\gamma = ((\alpha * v_\beta + \beta) * v_\gamma + \gamma) * o^3$

9: $\alpha\gamma\beta = ((\alpha * v_\beta + \beta) * v_\gamma + \gamma) * o^3$

10: $\beta\alpha\gamma = ((\alpha * v_\beta + \beta) * v_\gamma + \gamma) * o^3$

11: $\beta\gamma\alpha = ((\alpha * v_\beta + \beta) * v_\gamma + \gamma) * o^3$

12: $\gamma\beta\alpha = ((\alpha * v_\beta + \beta) * v_\gamma + \gamma) * o^3$

13: $\gamma\alpha\beta = ((\alpha * v_\beta + \beta) * v_\gamma + \gamma) * o^3$

14: ▷ Set up occupied indices

15: $ijk = (i * o + j) * o + k$

16: $ikj = (i * o + k) * o + j$

17: $jik = (j * o + i) * o + k$

18: $jki = (j * o + k) * o + i$

19: $kji = (k * o + j) * o + i$

20: $kij = (k * o + i) * o + j$

21: ▷ Bounds checking

22: **if** $i < o$ && $j < o$ && $k < o$ **then**

23: Build $D_{ijk}^{\alpha\beta\gamma} = \epsilon_i + \epsilon_j + \epsilon_k$ - $(\epsilon_a + \epsilon_b + \epsilon_c)$ ▷ Equation 4

24: Build $V_{ijk}^{\alpha\beta\gamma} = t_i^\alpha\ G_{jk}^{\beta\gamma} + t_j^\beta\ G_{ik}^{\alpha\gamma} + t_k^\gamma\ G_{ij}^{\alpha\beta}$ ▷ Equation 3 ▷ Construct $w_{ijk\alpha\beta\gamma}$ contributions

25: $wijk\alpha\beta\gamma = W_{ijk}^{\alpha\beta\gamma}[ijk + \alpha\beta\gamma] + W_{ijk}^{\alpha\gamma\beta}[ikj + \alpha\gamma\beta] + W_{ijk}^{\beta\alpha\gamma}[jik + \beta\alpha\gamma] + W_{ijk}^{\beta\gamma\alpha}[jki + \beta\gamma\alpha] + W_{ijk}^{\gamma\beta\alpha}[kji + \gamma\beta\alpha] + W_{ijk}^{\gamma\alpha\beta}[kij + \gamma\alpha\beta]$

26: $wjki\alpha\beta\gamma = W_{ijk}^{\alpha\beta\gamma}[jki + \alpha\beta\gamma] + W_{ijk}^{\alpha\gamma\beta}[jik + \alpha\gamma\beta] + W_{ijk}^{\beta\alpha\gamma}[kji + \beta\alpha\gamma] + W_{ijk}^{\beta\gamma\alpha}[kij + \beta\gamma\alpha] + W_{ijk}^{\gamma\beta\alpha}[ikj + \gamma\beta\alpha] + W_{ijk}^{\gamma\alpha\beta}[ijk + \gamma\alpha\beta]$

27: $wkij\alpha\beta\gamma = W_{ijk}^{\alpha\beta\gamma}[kij + \alpha\beta\gamma] + W_{ijk}^{\alpha\gamma\beta}[kji + \alpha\gamma\beta] + W_{ijk}^{\beta\alpha\gamma}[ikj + \beta\alpha\gamma] + W_{ijk}^{\beta\gamma\alpha}[ijk + \beta\gamma\alpha] + W_{ijk}^{\gamma\beta\alpha}[jik + \gamma\beta\alpha] + W_{ijk}^{\gamma\alpha\beta}[jki + \gamma\alpha\beta]$

28: $wjik\alpha\beta\gamma = W_{ijk}^{\alpha\beta\gamma}[jik + \alpha\beta\gamma] + W_{ijk}^{\alpha\gamma\beta}[jki + \alpha\gamma\beta] + W_{ijk}^{\beta\alpha\gamma}[ijk + \beta\alpha\gamma] + W_{ijk}^{\beta\gamma\alpha}[ikj + \beta\gamma\alpha] + W_{ijk}^{\gamma\beta\alpha}[kij + \gamma\beta\alpha] + W_{ijk}^{\gamma\alpha\beta}[kji + \gamma\alpha\beta]$

29: $wikj\alpha\beta\gamma = W_{ijk}^{\alpha\beta\gamma}[ikj + \alpha\beta\gamma] + W_{ijk}^{\alpha\gamma\beta}[ijk + \alpha\gamma\beta] + W_{ijk}^{\beta\alpha\gamma}[kij + \beta\alpha\gamma] + W_{ijk}^{\beta\gamma\alpha}[kji + \beta\gamma\alpha] + W_{ijk}^{\gamma\beta\alpha}[jki + \gamma\beta\alpha] + W_{ijk}^{\gamma\alpha\beta}[jik + \gamma\alpha\beta]$

30: $wkji\alpha\beta\gamma = W_{ijk}^{\alpha\beta\gamma}[kji + \alpha\beta\gamma] + W_{ijk}^{\alpha\gamma\beta}[kij + \alpha\gamma\beta] + W_{ijk}^{\beta\alpha\gamma}[jki + \beta\alpha\gamma] + W_{ijk}^{\beta\gamma\alpha}[jik + \beta\gamma\alpha] + W_{ijk}^{\gamma\beta\alpha}[ijk + \gamma\beta\alpha] + W_{ijk}^{\gamma\alpha\beta}[ikj + \gamma\alpha\beta]$

31: $sum_{ijk}[ijk] += (wijk\alpha\beta\gamma + V_{ijk}^{\alpha\beta\gamma}) * (4wijk\alpha\beta\gamma + wjki\alpha\beta\gamma + wkij\alpha\beta\gamma - 2wjik\alpha\beta\gamma - 2wikj\alpha\beta\gamma - 2wkji\alpha\beta\gamma)$ (Equation 1)

32: **end if**

**Figure 4**. Pseudocode of the energy evaluation kernel. As before, arrays stored in device (GPU) memory are shown in green text. The position of each thread in the block determines the index set it is assigned (lines 4-6). The compound indices needed for fetching the energy contributions are set up in lines 8-13 (virtual) and lines 15-20 (occupied). The energy denominators $D_{ijk}^{\alpha\beta\gamma}$ and $V_{ijk}^{\alpha\beta\gamma}$ tensor are built in lines 23 and 24. Next, the contributions of the various permutations of $W_{ijk}^{\alpha\beta\gamma}$ are accumulated in lines 25-30, and the final energy contribution is computed in line 31.

Note that an additional multiplication by a scaling factor representing the Kronecker deltas in Equation 1 is not shown here.

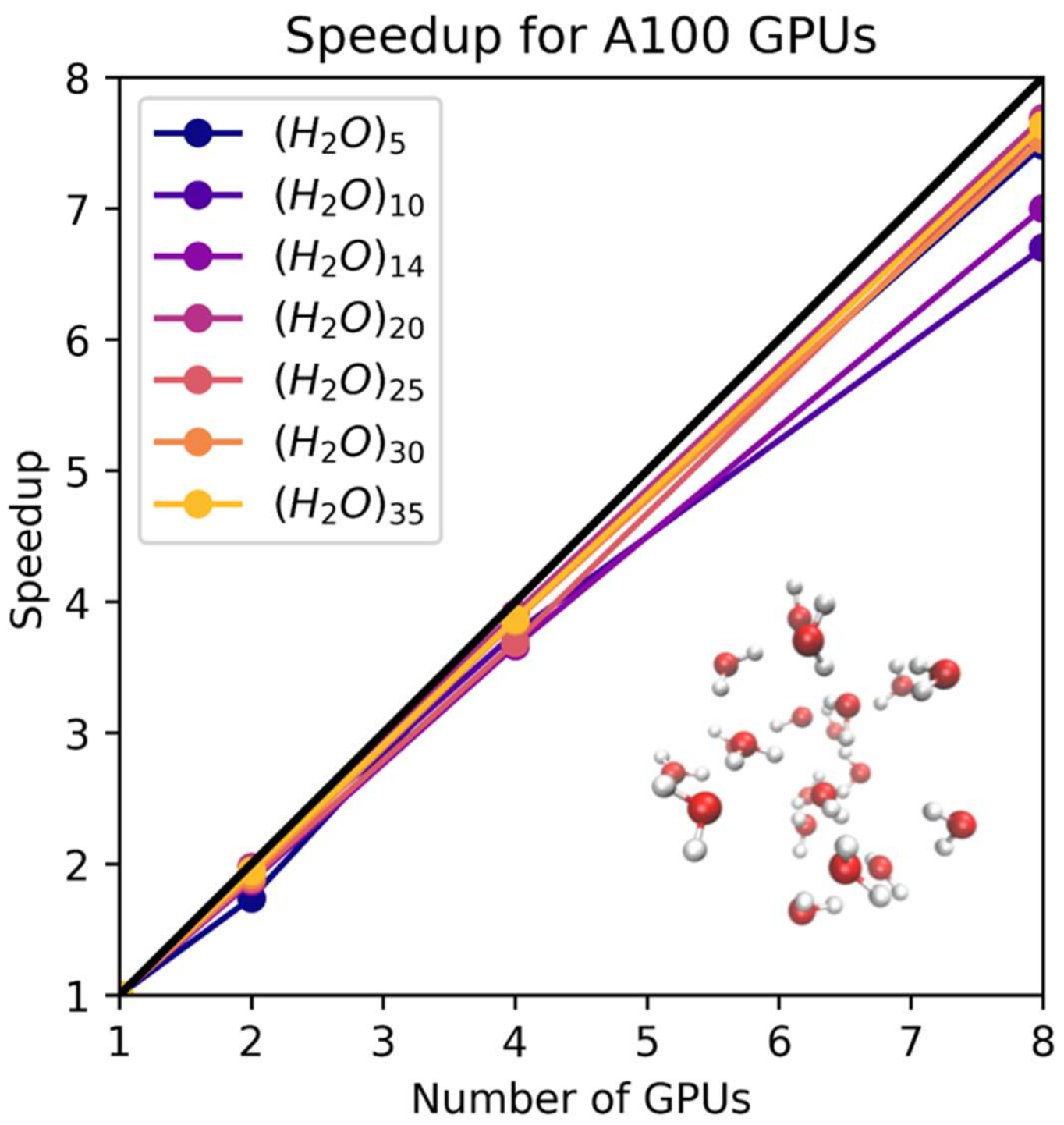

**Figure 5**: Strong scaling of (T) calculation in the cc-pVDZ basis set for 5 through 35 water molecules. The calculations were performed on a node containing 8 A100 GPUs. The speedup for N GPUs is given by the ratio of the time taken by 1 GPU to the time taken by N.

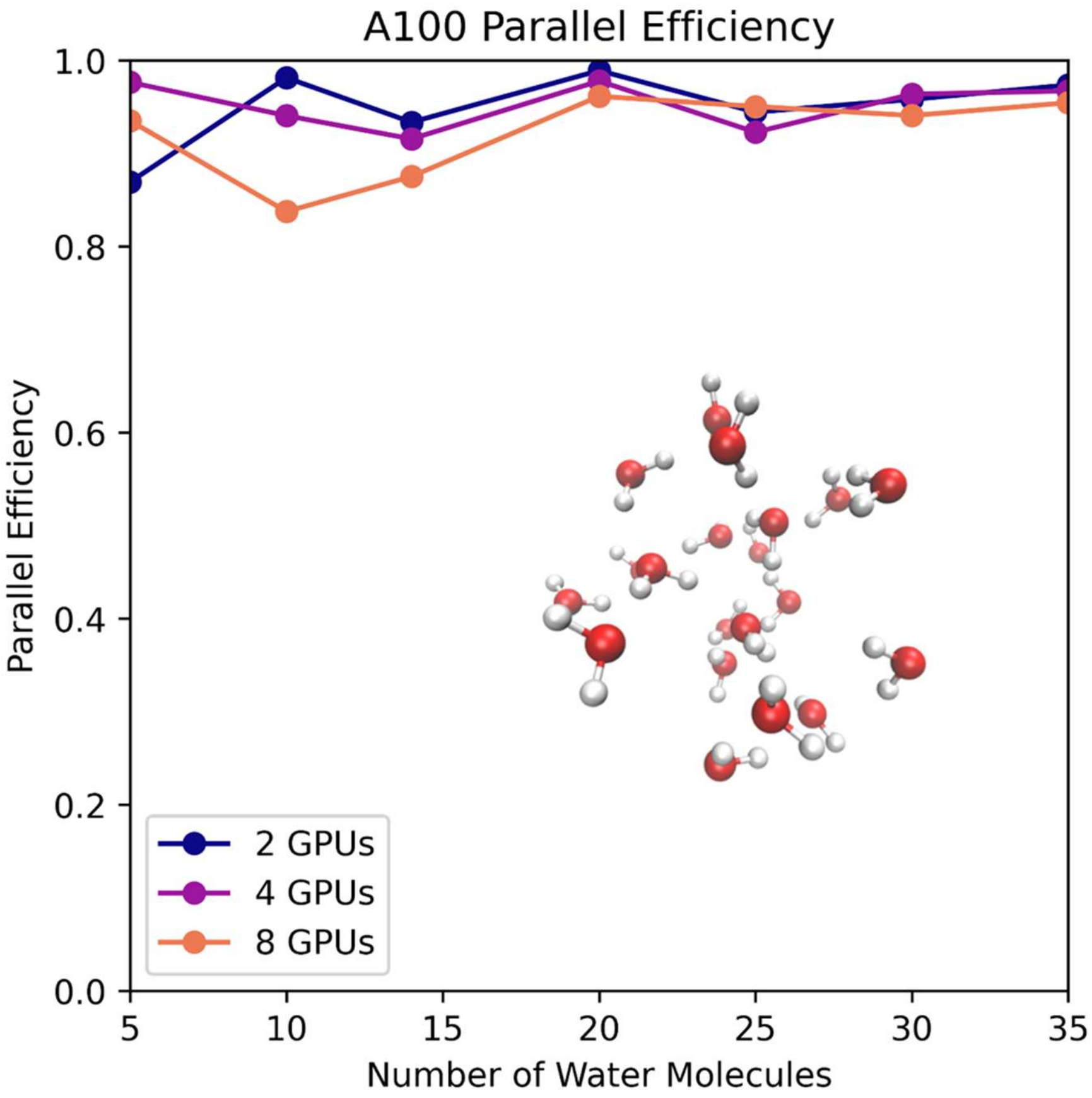

**Figure 6.** Parallel efficiency for the (T) portion of the CCSD(T) calculation, given as the time taken by 1 GPU divided by N times the time for N GPUs.

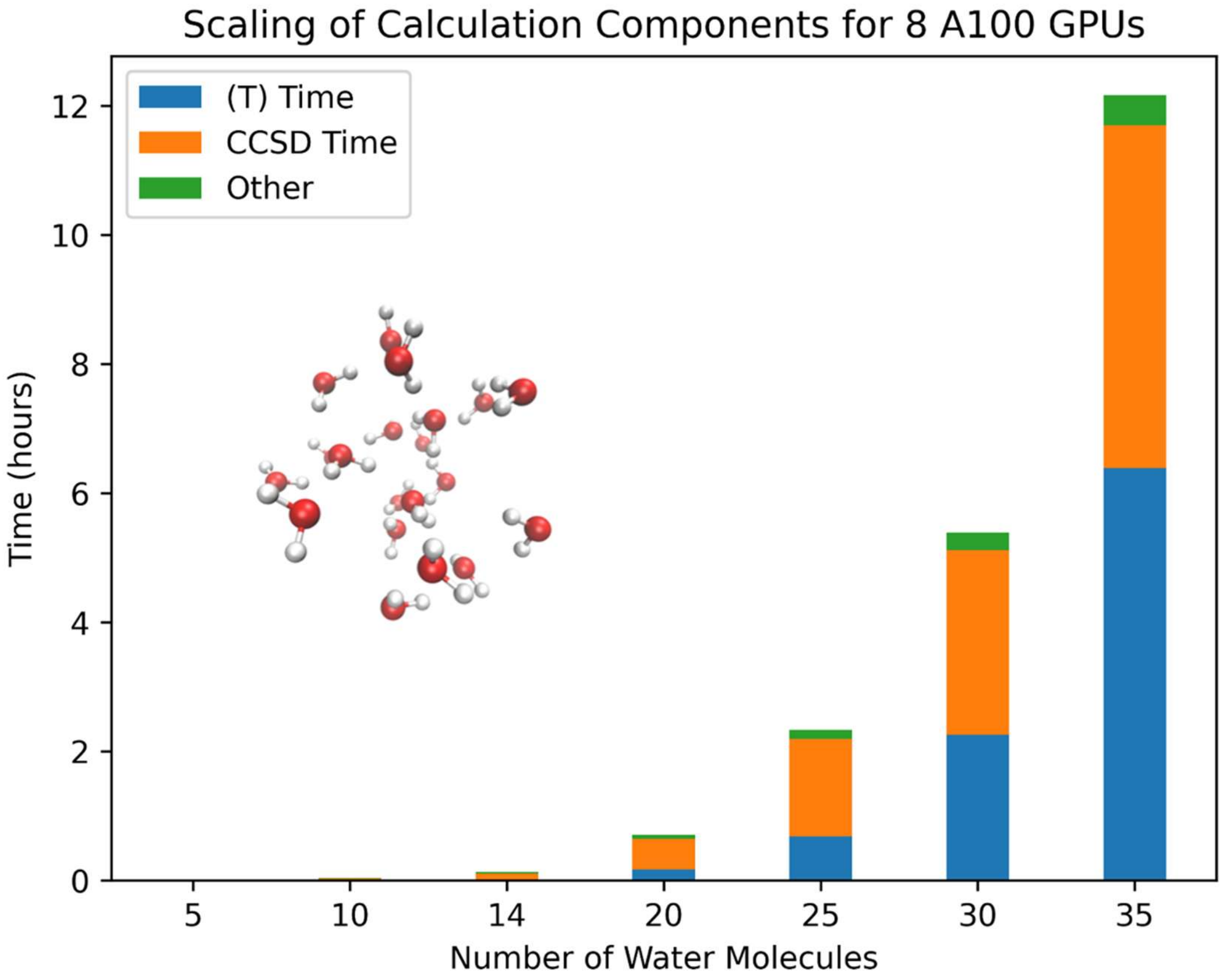

**Figure 7.** Breakdown of computational time spent of various portions of the total CCSD(T) calculation on 8 A100 GPUs. Even for the largest system the (T) time does not overtake the CCSD time.

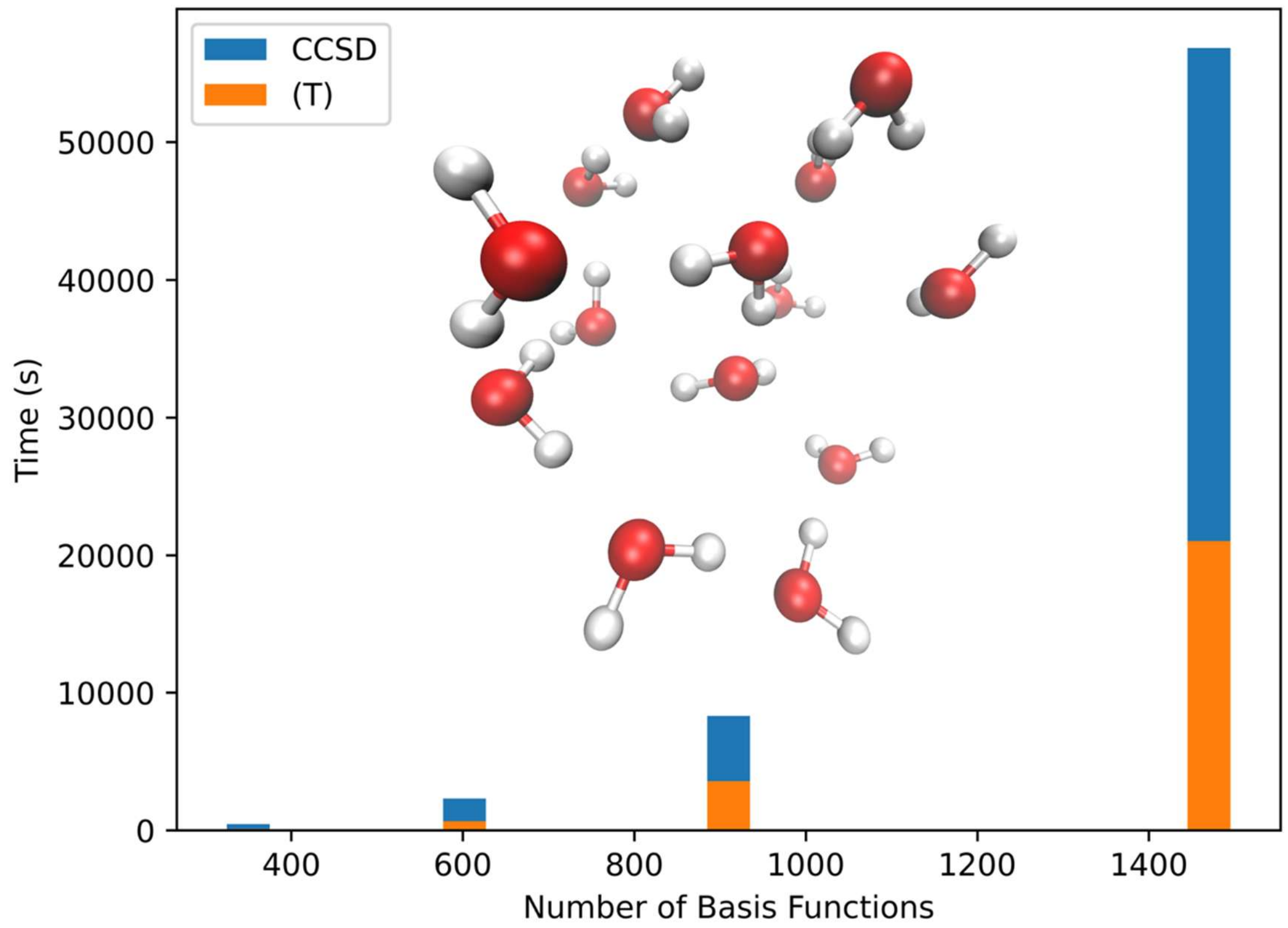

**Figure 8**: Timings for CCSD(T) calculation of 14 water molecules using cc-pVDZ, aug-cc-pVDZ, cc-pVTZ, and aug-cc-pVTZ basis sets (with 350, 602, 910, and 1470 basis functions, respectively) running on 8 A100 GPUs.

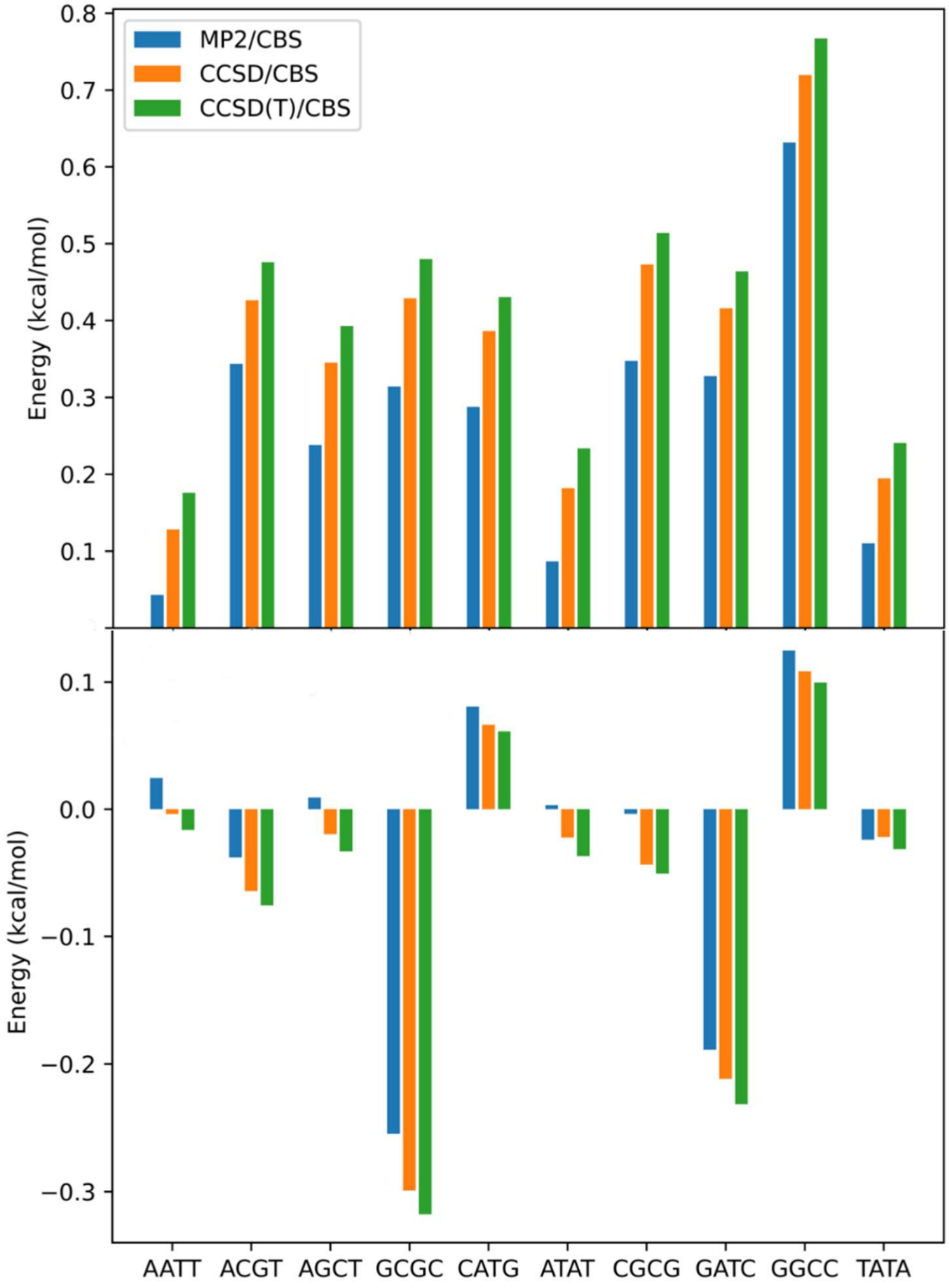

**Figure 9.** The average non-additive 3-body interaction energy (upper panel) and the non-additive 4-body interaction energy (lower panel) for all 10 unique DNA configurations.

**Table 1.** Timing comparison for GPU-accelerated CCSD(T) of $(H_2O)_{14}$ in the cc-pVDZ basis set using MPQC and the implementation described in the text (TeraChem). Details about the hardware employed in the comparisons can be found in Table S1 in the SI.

| Software | GPUs | (T) Time (min) |
|---|---|---|
| MPQC[42] | 1 P100 | 67 |
| MPQC[42] | 2 P100 | 39 |
| TeraChem | 1 P100 | 34.95 |
| TeraChem | 2 P100 | 18.53 |

**Table 2.** Comparison of time and node hours required to calculate (T) energy correction with several multithreaded implementations of CCSD(T). Details about the hardware employed in the comparisons can be found in Table S1 of the SI.

| Software | ERIs | Nodes | Total Cores | GPUs | (T) Time (min) | Node-hours |
|---|---|---|---|---|---|---|
| | | | *Uracil, cc-pVDZ-F12, 12 atoms, 276 AOs* | | | |
| MPQC[42] | DF | 1 | 2 | 0 | 16 | 0.27 |
| TeraChem | CD | 1 | 1 | 1 A100 | 1.2 | 0.02 |
| | | | *Pentacene, cc-pVDZ, 36 atoms, 378 AOs* | | | |
| MPQC[42] | DF | 1 | 2 | 0 | 507 | 8.45 |
| TeraChem | CD | 1 | 1 | 1 A100 | 9.5 | 0.16 |
| | | | *$(H_2O)_{10}$, cc-pVDZ, 30 atoms, 240 AOs* | | | |
| MRCC[50] | DF | 1 | 16 | 0 | 13.9 | 0.23 |
| ORCA[50] | DF | 1 | 16 | 0 | 16.4 | 0.27 |
| Psi4[50] | DF | 1 | 16 | 0 | 21.8 | 0.36 |
| GAMESS[49] | DF | 1 | 16 | 0 | 18.2 | 0.30 |
| TeraChem | CD | 1 | 1 | 1 A100 | 1.4 | 0.02 |

**Table 3.** Comparison of time and node hours required to calculate (T) energy correction with several distributed memory parallel implementations of CCSD(T). Details about the hardware employed in the comparisons can be found in Table S1 of the SI.

| Software | ERIs | Nodes | Total Cores | GPUs | (T) Time (min) | Node-hours |
|---|---|---|---|---|---|---|
| | | | *$(H_2O)_{14}$, cc-pVDZ, 42 Atoms, 336 AOs* | | | |
| MRCC[50] | DF | 8 | 128 | 0 | 22 | 2.93 |
| TeraChem | CD | 1 | 8 | 8 A100 | 1.16 | 0.02 |
| | | | *Coronene, cc-pVDZ, 36 Atoms, 396 AOs* | | | |
| GAMESS[49] | DF | 128 | 8192 | 0 | 6 | 12.8 |
| TeraChem | CD | 1 | 1 | 1 A100 | 12.4 | 0.21 |
| TeraChem | CD | 1 | 8 | 8 A100 | 1.7 | 0.03 |
| | | | *$(H_2O)_{20}$, cc-pVDZ, 60 Atoms, 480 AOs* | | | |
| FHI-aims[53] | DF | 128 | 2560 | 0 | 183 | 390.4 |
| TeraChem | CD | 1 | 8 | 8 A100 | 10.17 | 0.1695 |
| | | | *Uracil dimer, cc-pVDZ-F12, 24 Atoms, 552 AOs* | | | |
| MPQC[42] | DF | 8 | 16 | 0 | 152 | 20.27 |
| TeraChem | CD | 1 | 8 | 8 A100s | 3.54 | 0.059 |
| | | | *Pentacene dimer, cc-pVDZ, 72 Atoms, 756 AOs* | | | |
| MPQC[42] | DF | 32 | 512 | 0 | 1485 | 792 |
| TeraChem | CD | 1 | 8 | 8 A100s | 102 | 1.70 |
| | | | *GC-dDMP-B, 6-311++G**, 63 Atoms, 1042 AOs* | | | |
| MPQC[42] | DF | 64 | 1024 | 0 | 2844 | 3034 |
| NWChem[46] | Exact | 20000 | 40000 | 0 | 84 | 28000 |
| TeraChem | CD | 1 | 8 | 8 A100 | 477 | 7.95 |

**Table 4.** DNA Pair Stacking Energy (kcal/mol) computed at various levels of theory, as defined in Eq, 9. "Approx1-CCSD(T)" results are corrected[72] data from Sponer, et al,[69] where individual base-pair interactions were estimated by MP2/aug-cc-pVDZ/aug-cc-pVTZ extrapolations plus ΔCCSD(T)/6-31G*(0.25) corrections and the sum of the pairwise interactions was corrected with a four-body MP2/aug-cc-pVDZ correction. "Approx2-CCSD(T)" is data from Hill, et al.,[72] where values were estimated from an aug-cc-pVTZ/aug-cc-pVQZ extrapolation of DF-LMP2 energies plus the ΔCCSD(T)/6-31G*(0.25) corrections from Sponer, et al.[69] "SAPT0/jaDZ" are results from Parker, et al.,[68] obtained using symmetry-adapted perturbation theory in a tuned basis set. "Full CCSD(T)" are results obtained in this work using full CCSD(T)/aug-cc-pVDZ energies for the tetramers along with MP2/aug-cc-pV5Z energies to estimate the CBS limit of CCSD(T) using Eqs. 7 and 8.

| Structure | Approx1-CCSD(T) | Approx2-CCSD(T) | SAPT0/jaDZ | Full CCSD(T) |
|---|---|---|---|---|
| AATT | -13.1 | -11.98 | -12.1 | -13.49 |
| ACGT | -13.4 | -12.01 | -11.29 | -13.48 |
| AGCT | -13.5 | -12.39 | -11.2 | -13.14 |
| GCGC | -15.4 | -14.7 | -14.48 | -16.55 |
| CATG | -15.1 | -13.96 | -13.63 | -14.38 |
| ATAT | -13.3 | -11.99 | -10.87 | -12.82 |
| CGCG | -17.3 | -16.16 | -15.69 | -16.46 |
| GATC | -12.9 | -11.26 | -10.22 | -12.21 |
| GGCC | -11.5 | -10.29 | -9.32 | -11.53 |
| TATA | -12.8 | | -11.92 | -13.03 |

## References

1. Čížek, J., On the Correlation Problem in Atomic and Molecular Systems. Calculation of Wavefunction Components in Ursell-Type Expansion Using Quantum-Field Theoretical Methods. *J. Chem. Phys.* **1966,** *45*, 4256.
2. Purvis, G. D.; Bartlett, R. J., A full coupled-cluster singles and doubles model: The inclusion of disconnected triples. *J. Chem. Phys.* **1982,** *76*, 1910.
3. Raghavachari, K.; Trucks, G. W.; Pople, J. A.; Head-Gordon, M., A fifth-order perturbation comparison of electron correlation theories. *Chem. Phys. Lett.* **1989,** *157*, 479.
4. Kaka, K. S.; Castet, F.; Champagne, B., On the third-order nonlinear optical responses of *cis* and *trans* stilbenes – a quantum chemistry investigation. *Phys. Chem. Chem. Phys.* **2024**, 14808.
5. Kaka, K. S.; Beaujean, P.; Castet, F.; Champagne, B., A quantum chemical investigation of the second hyperpolarizability of p-nitroaniline. *J. Chem. Phys.* **2023,** *159*, 114104.
6. Sinnokrot, M. O.; Sherrill, C. D., High-Accuracy Quantum Mechanical Studies of π−π Interactions in Benzene Dimers. *J. Phys. Chem. A* **2006,** *110*, 10656.
7. Tsuzuki, S.; Uchimaru, T.; Matsumura, K.; Mikami, M.; Tanabe, K., Effects of the higher electron correlation correction on the calculated intermolecular interaction energies of benzene and naphthalene dimers: comparison between MP2 and CCSD(T) calculations. *Chem. Phys. Lett.* **2000,** *319*, 547.
8. Riley, K. E.; Pitoňák, M.; Jurečka, P.; Hobza, P., Stabilization and Structure Calculations for Noncovalent Interactions in Extended Molecular Systems Based on Wave Function and Density Functional Theories. *Chem. Rev.* **2010,** *110*, 5023.
9. Saebo, S.; Pulay, P., Local Treatment of Electron Correlation. *Ann. Rev. Phys. Chem.* **1993,** *44*, 213.
10. Pulay, P., Localizability of dynamic electron correlation. *Chem. Phys. Lett.* **1983,** *100*, 151.
11. Laidig, W. D.; Purvis, G. D.; Bartlett, R. J., Localized orbitals in the coupled cluster singles and doubles model. *Int. J. Quantum Chem.* **2009,** *22*, 561.
12. Schütz, M.; Werner, H.-J., Low-order scaling local electron correlation methods. IV. Linear scaling local coupled-cluster (LCCSD). *J. Chem. Phys.* **2001,** *114*, 661.
13. Bensberg, M.; Neugebauer, J., Orbital pair selection for relative energies in the domain-based local pair natural orbital coupled-cluster method. *J. Chem. Phys.* **2022,** *157*, 064102.
14. Davidson, E. R., Selection of the Proper Canonical Roothaan-Hartree-Fock Orbitals for Particular Applications. I. Theory. *J. Chem. Phys.* **1972,** *57*, 1999.
15. Nagy, P. R.; Kállay, M., Approaching the Basis Set Limit of CCSD(T) Energies for Large Molecules with Local Natural Orbital Coupled-Cluster Methods. *J. Chem. Theory Comput.* **2019,** *15*, 5275.
16. Liakos, D. G.; Guo, Y.; Neese, F., Comprehensive Benchmark Results for the Domain Based Local Pair Natural Orbital Coupled Cluster Method (DLPNO-CCSD(T)) for Closed- and Open-Shell Systems. *J. Phys. Chem. A* **2020,** *124*, 90.
17. Neese, F.; Hansen, A.; Liakos, D. G., Efficient and accurate approximations to the local coupled cluster singles doubles method using a truncated pair natural orbital basis. *J. Chem. Phys.* **2009,** *131*, 064103.

18. Saitow, M.; Becker, U.; Riplinger, C.; Valeev, E. F.; Neese, F., A new near-linear scaling, efficient and accurate, open-shell domain-based local pair natural orbital coupled cluster singles and doubles theory. *J. Chem. Phys.* **2017,** *146*, 164105.
19. Lesiuk, M., Implementation of the Coupled-Cluster Method with Single, Double, and Triple Excitations using Tensor Decompositions. *J. Chem. Theory Comput.* **2020,** *16*, 453.
20. Lesiuk, M., Quintic-scaling rank-reduced coupled cluster theory with single and double excitations. *J. Chem. Phys.* **2022,** *156*, 064103.
21. Lesiuk, M., When Gold Is Not Enough: Platinum Standard of Quantum Chemistry with $N^7$ Cost. *J. Chem. Theory Comput.* **2022,** *18*, 6537.
22. Hohenstein, E. G.; Parrish, R. M.; Martínez, T. J., Tensor hypercontraction density fitting. I. Quartic scaling second- and third-order Møller-Plesset perturbation theory. *J. Chem. Phys.* **2012,** *137*, 044103.
23. Parrish, R. M.; Hohenstein, E. G.; Martínez, T. J.; Sherrill, C. D., Tensor hypercontraction. II. Least-squares renormalization. *J. Chem. Phys.* **2012,** *137*, 224106.
24. Hohenstein, E. G.; Parrish, R. M.; Sherrill, C. D.; Martínez, T. J., Communication: Tensor hypercontraction. III. Least-squares tensor hypercontraction for the determination of correlated wavefunctions. *J. Chem. Phys.* **2012,** *137*, 221101.
25. Hohenstein, E. G.; Kokkila, S. I. L.; Parrish, R. M.; Martínez, T. J., Quartic scaling second-order approximate coupled cluster singles and doubles via tensor hypercontraction: THC-CC2. *J. Chem. Phys.* **2013,** *138*, 124111.
26. Parrish, R. M.; Hohenstein, E. G.; Schunck, N. F.; Sherrill, C. D.; Martínez, T. J., Exact Tensor Hypercontraction: A Universal Technique for the Resolution of Matrix Elements of Local Finite-Range N -Body Potentials in Many-Body Quantum Problems. *Phys. Rev. Lett.* **2013,** *111*, 132505.
27. Parrish, R. M.; Hohenstein, E. G.; Martínez, T. J.; Sherrill, C. D., Discrete variable representation in electronic structure theory: Quadrature grids for least-squares tensor hypercontraction. *J. Chem. Phys.* **2013,** *138*, 194107.
28. Hohenstein, E. G.; Kokkila, S. I. L.; Parrish, R. M.; Martínez, T. J., Tensor Hypercontraction Equation-of-Motion Second-Order Approximate Coupled Cluster: Electronic Excitation Energies in O(N4) Time. *J. Phys. Chem. B* **2013,** *117*, 12972.
29. Parrish, R. M.; Sherrill, C. D.; Hohenstein, E. G.; Kokkila, S. I. L.; Martínez, T. J., Communication: Acceleration of coupled cluster singles and doubles via orbital-weighted least-squares tensor hypercontraction. *J. Chem. Phys.* **2014,** *140*, 181102.
30. Kokkila Schumacher, S. I. L.; Hohenstein, E. G.; Parrish, R. M.; Wang, L.-P.; Martínez, T. J., Tensor Hypercontraction Second-Order Møller–Plesset Perturbation Theory: Grid Optimization and Reaction Energies. *J. Chem. Theory Comput.* **2015,** *11*, 3042.
31. Parrish, R. M.; Zhao, Y.; Hohenstein, E. G.; Martínez, T. J., Rank reduced coupled cluster theory. I. Ground state energies and wavefunctions. *J. Chem. Phys.* **2019,** *150*, 164118.
32. Hohenstein, E. G.; Zhao, Y.; Parrish, R. M.; Martínez, T. J., Rank reduced coupled cluster theory. II. Equation-of-motion coupled-cluster singles and doubles. *J. Chem. Phys.* **2019,** *151*, 164121.
33. Hohenstein, E. G.; Martínez, T. J., GPU acceleration of rank-reduced coupled-cluster singles and doubles. *J. Chem. Phys.* **2021,** *155*, 184110.
34. Hohenstein, E. G.; Fales, B. S.; Parrish, R. M.; Martínez, T. J., Rank-reduced coupled-cluster. III. Tensor hypercontraction of the doubles amplitudes. *J. Chem. Phys.* **2022,** *156*, 054102.

35. Ruth, M.; Gerbig, D.; Schreiner, P. R., Machine Learning of Coupled Cluster (T)-Energy Corrections via Delta (Δ)-Learning. *J. Chem. Theory Comput.* **2022,** *18*, 4846.
36. Ruth, M.; Gerbig, D.; Schreiner, P. R., Machine Learning for Bridging the Gap between Density Functional Theory and Coupled Cluster Energies. *J. Chem. Theory Comput.* **2023,** *19*, 4912.
37. Sparrow, Z. M.; Ernst, B. G.; Joo, P. T.; Lao, K. U.; DiStasio, R. A., NENCI-2021. I. A large benchmark database of non-equilibrium non-covalent interactions emphasizing close intermolecular contacts. *J. Chem. Phys.* **2021,** *155*, 184303.
38. Kaser, S.; Meuwly, M., Transfer-learned potential energy surfaces: Toward microsecond-scale molecular dynamics simulations in the gas phase at CCSD(T) quality. *J. Chem. Phys.* **2023,** *158*, 214301.
39. Daru, J.; Forbert, H.; Behler, J.; Marx, D., Coupled Cluster Molecular Dynamics of Condensed Pahse Systems Enabled by Machine Learning Potentials: Liquid Water Benchmark. *Phys. Rev. Lett.* **2022,** *129*, 226001.
40. Messerly, M.; Matin, S.; Allen, A. E.; Nebgen, B.; Barros, K.; Smith, J. S.; Lubbers, N.; Messerly, R., Multi-fidelity learning for interatomic potentials: low-level forces and high-level energies are all you need. *Mach. Learn. Sci. Tech.* **2025,** *6*, 035066.
41. Watts, J. D., Parallel algorithms for coupled-cluster methods. *Par. Comp.* **2000,** *26*, 857.
42. Peng, C.; Calvin, J. A.; Valeev, E. F., Coupled-cluster singles, doubles and perturbative triples with density fitting approximation for massively parallel heterogeneous platforms. *Int. J. Quantum Chem.* **2019,** *119*, e25894.
43. Calvin, J. A.; Peng, C.; Rishi, V.; Kumar, A.; Valeev, E. F., Many-Body Quantum Chemistry on Massively Parallel Computers. *Chem. Rev.* **2021,** *121*, 1203.
44. Rendell, A. P.; Lee, T. J.; Komornicki, A.; Wilson, S., Evaluation of the contribution from triply excited intermediates to the fourth-order perturbation theory energy on Intel distributed memory supercomputers. *Theo. Chim. Acta* **1993,** *84*, 271.
45. Kobayashi, R.; Rendell, A. P., A direct coupled cluster algorithm for massively parallel computers. *Chem. Phys. Lett.* **1997,** *265*, 1.
46. Anisimov, V. M.; Bauer, G. H.; Chadalavada, K.; Olson, R. M.; Glenski, J. W.; Kramer, W. T. C.; Aprà, E.; Kowalski, K., Optimization of the Coupled Cluster Implementation in NWChem on Petascale Parallel Architectures. *J. Chem. Theory Comput.* **2014,** *10*, 4307.
47. Asadchev, A.; Gordon, M. S., Fast and Flexible Coupled Cluster Implementation. *J. Chem. Theory Comput.* **2013,** *9*, 3385.
48. Janowski, T.; Pulay, P., Efficient Parallel Implementation of the CCSD External Exchange Operator and the Perturbative Triples (T) Energy Calculation. *J. Chem. Theory Comput.* **2008,** *4*, 1585.
49. Datta, D.; Gordon, M. S., A Massively Parallel Implementation of the CCSD(T) Method Using the Resolution-of-the-Identity Approximation and a Hybrid Distributed/Shared Memory Parallelization Model. *J. Chem. Theory Comput.* **2021,** *17*, 4799.
50. Gyevi-Nagy, L.; Kállay, M.; Nagy, P. R., Integral-Direct and Parallel Implementation of the CCSD(T) Method: Algorithmic Developments and Large-Scale Applications. *J. Chem. Theory Comput.* **2020,** *16*, 366.
51. Mester, D.; Nagy, P. R.; Csoka, J.; Gyevi-Nagy, L.; Szabo, P. B.; Horvath, R. A.; Petrov, K.; Hegely, B.; Ladoczki, B.; Samu, G., et al., Overview of Developments in the MRCC Program System. *J. Phys. Chem. A* **2025,** *129*, 2086.

52. Blum, V.; Gehrke, R.; Hanke, F.; Havu, P.; Havu, V.; Ren, X.; Reuter, K.; Scheffler, M., Ab initio molecular simulations with numeric atom-centered orbitals. *Comp. Phys. Comm.* **2009,** *180*, 2175.
53. Shen, T.; Zhu, Z.; Zhang, I. Y.; Scheffler, M., Massive-Parallel Implementation of the Resolution-of-Identity Coupled-Cluster Approaches in the Numeric Atom-Centered Orbital Framework for Molecular Systems. *J. Chem. Theory Comput.* **2019,** *15*, 4721.
54. Hirata, S., Tensor Contraction Engine: Abstraction and Automated Parallel Implementation of Configuration-Interaction, Coupled-Cluster, and Many-Body Perturbation Theories. *J. Phys. Chem. A* **2003,** *107*, 9887.
55. Lotrich, V.; Flocke, N.; Ponton, M.; Yau, A. D.; Perera, A.; Deumens, E.; Bartlett, R. J., Parallel implementation of electronic structure energy, gradient, and Hessian calculations. *J. Chem. Phys.* **2008,** *128*, 194104.
56. Datta, D.; Gordon, M. S., Accelerating Coupled-Cluster Calculations with GPUs: An Implementation of the Density-Fitted CCSD(T) Approach for Heterogeneous Computing Architectures Using OpenMP Directives. *J. Chem. Theory Comput.* **2023,** *19*, 7640.
57. Kim, J.; Panyala, A.; Peng, B.; Kowalski, K.; Sadayappan, P.; Krishnamoorthy, S., Scalable Heterogeneous Execution of a Coupled-Cluster Model with Perturbative Triples. In *SC20: International Conference for High Performance Computing, Networking, Storage and Analysis*, 2020; pp 1.
58. Ufimtsev, I. S.; Martínez, T. J., Quantum Chemistry on Graphical Processing Units. 1. Strategies for Two-Electron Integral Evaluation. *J. Chem. Theory Comput.* **2008,** *4*, 222.
59. Ufimtsev, I. S.; Martinez, T. J., Quantum Chemistry on Graphical Processing Units. 3. Analytical Energy Gradients, Geometry Optimization, and First Principles Molecular Dynamics. *J. Chem. Theory Comput.* **2009,** *5*, 2619.
60. Ufimtsev, I. S.; Martinez, T. J., Quantum Chemistry on Graphical Processing Units. 2. Direct Self-Consistent-Field Implementation. *J. Chem. Theory Comput.* **2009,** *5*, 1004.
61. Titov, A. V.; Ufimtsev, I. S.; Luehr, N.; Martinez, T. J., Generating Efficient Quantum Chemistry Codes for Novel Architectures. *J. Chem. Theory Comput.* **2013,** *9*, 213.
62. Fales, B. S.; Curtis, E. R.; Johnson, K. G.; Lahana, D.; Seritan, S.; Wang, Y.; Weir, H.; Martínez, T. J.; Hohenstein, E. G., Performance of Coupled-Cluster Singles and Doubles on Modern Stream Processing Architectures. *J. Chem. Theory Comput.* **2020,** *16*, 4021.
63. Beebe, N. H. F.; Linderberg, J., Simplifications in the Generation and Transformation of Two-Electron Integrals in Molecular Calculations. *Int. J. Quantum Chem.* **1977,** *12*, 683.
64. Feyereisen, M.; Fitzgerald, G.; Komornicki, A., Use of Approximate Integrals in Ab Initio Theory. An Application in MP2 Energy Calculations. *Chem. Phys. Lett.* **1993,** *208*, 359.
65. Koch, H.; Sanchez De Meras, A.; Pedersen, T. B., Reduced Scaling in Electronic Structure Calculations Using Cholesky Decomposition. *J. Chem. Phys.* **2003,** *118*, 9481.
66. Kendall, R. A.; Dunning, T. H.; Harrison, R. J., Electron Affinities of the First-Row Atoms Revisited. Systematic Basis Sets and Wave Functions. *J. Chem. Phys.* **1992,** *96*, 6796.
67. Dunning, T. H., Gaussian Basis Sets for Use in Correlated Molecular Calculations. I. The Atoms Boron through Neon and Hydrogen. *J. Chem. Phys.* **1989,** *90*, 1007.
68. Parker, T. M.; Hohenstein, E. G.; Parrish, R. M.; Hud, N. V.; Sherrill, C. D., Quantum-Mechanical Analysis of the Energetic Contributions to π Stacking in Nucleic Acids versus Rise, Twist, and Slide. *J. Amer. Chem. Soc.* **2013,** *135*, 1306.

69. Šponer, J.; Jurečka, P.; Marchan, I.; Luque, F. J.; Orozco, M.; Hobza, P., Nature of Base Stacking: Reference Quantum-Chemical Stacking Energies in Ten Unique B-DNA Base-Pair Steps. *Chem. Eur. J.* **2006,** *12*, 2854.

70. Halkier, A.; Helgaker, T.; Jørgensen, P.; Klopper, W.; Koch, H.; Olsen, J.; Wilson, A. K., Basis-set convergence in correlated calculations on Ne, N2, and H2O. *Chem. Phys. Lett.* **1998,** *286*, 243.

71. Smith, D. G. A.; Burns, L. A.; Simmonett, A. C.; Parrish, R. M.; Schieber, M. C.; Galvelis, R.; Kraus, P.; Kruse, H.; Di Remigio, R.; Alenaizan, A., et al., PSI4 1.4: Open-source software for high-throughput quantum chemistry. *J. Chem. Phys.* **2020,** *152*, 184108.

72. Hill, J. G.; Platts, J. A., Calculating stacking interactions in nucleic acid base-pair steps using spin-component scaling and local second order Møller–Plesset perturbation theory. *Phys. Chem. Chem. Phys.* **2008,** *10*, 2785.

73. Each of the 7763 CCSD(T) noncovalent interaction calculations requires three sub-calculations (dimer and two monomers). The largest atoms represented in these calculations are Br and are treated with the aug-cc-pVTZ basis set and the largest dimers have 16 heavy atoms. If we would assume that each heavy atom was Br, and that each calculation would be as large as the largest, we would need to do 7763 x 3 = 23,289 calculations with about 800 basis functions each. Based on our calculations, a node with 8 A100s would need ~2 node hours for the CCSD calculation and ~1 node hour for the (T) correction. This worst-case scenario would correspond to ~70,000 node hours to complete the dataset, which should represent a conservative upper bound.

**Supplemental Information for**
**Accelerating CCSD(T) on Graphical Processing Units (GPUs)**

O. Jonathan Fajen,[1,2,3*] Joseph E. Kelly,[1*] Edward G. Hohenstein,[1,2,3] Todd J. Martinez[1,2,3]

[1]Department of Chemistry, Stanford University, Stanford, California 94305, United States

[2]The PULSE Institute, Stanford University, Stanford, California 94305, United States

[3]SLAC National Accelerator Laboratory, Menlo Park, California 94024, United States

**Table S1. CCSD(T) timing info.** Here we collect descriptions of the CPU (and GPU if applicable) hardware used for the timings of various other CCSD(T) implementations quoted in the main text. Note that we make no effort to describe compilation environments. We emphasize that the variety in hardware employed by these studies makes a direct comparison of the reported wall times highly challenging. Rather, the presentation of competing implementations serves to provide context for the performance that one can expect from an efficient CCSD(T) code.

| Program | Hardware |
|---|---|
| MPQC[1] | For smaller systems: CPU- 2 Intel Xeon E5-2670 CPUs (332 GFLOPS peak double) and 64 GB RAM (For single node calculations, 1 MPI process and 64 threads; for multi-node 4 MPI process per node with 16 threads per proc)<br>GPU- 2 Intel Xeon E5-2680v4 CPUs (1075 GFLOPS peak double), 512 GB RAM and 2 NVIDIA P100 GPUs with 12 GB HBM2 memory.<br>For GC-dDMP: 64 nodes on BlueRidge cluster. |
| MRCC, ORCA, Psi4[2] | Each node has 2 Intel Xeon E5-2650 v2 CPUs (8 physical cores per CPU). 125 GB RAM (Peak 332.8 GFLOPS)<br>Section 6 calculations: Intel Xeon Platinum 8180M CPUs with 28 physical cores each (Peak 1523 GFLOPS) |
| GAMESS[3] | $(H_2O)_{10}$: each node has 2 12-core Intel Xeon E5-2695 v2 CPUs with 126 GB RAM<br>Coronene: ALCF system. Each node has 64-core Intel KNL 7230 compute nodes with 192 GB RAM |
| FHI-aims[4] | HYDRA/MPCDF system. 610 nodes each with 2 Intel Sandy Bridge Xeon E5-2670 CPUs and 8 cores per CPU with 64 GB RAM.<br>3500 nodes each with 2 Ivy Bridge E5-2680 CPUs with 10 physical cores per CPU and 64 GB RAM (although 20 Sandy and 200 Ivy nodes have 128 GB RAM). $(H_2O)_{10}$ was done on SB, $(H_2O)_{15}$ and $(H_2O)_{20}$ was performed on IB. |
| NWChem[5] | 20,000 nodes. Each node has 2 AMD 6276 Interlagos Processors (313.6 GFLOPS) and 64 GB (Blue Waters) |
| TeraChem | Timings for A100 and V100 GPUs were performed using a DGX node with 2 64-core AMD EPYC 7742 CPUs and 8 A100s with 40 GB of memory, or 8 V100s with 32 GB. The calculations on P100s were performed on a node with 2 16 GB P100 GPUs. |

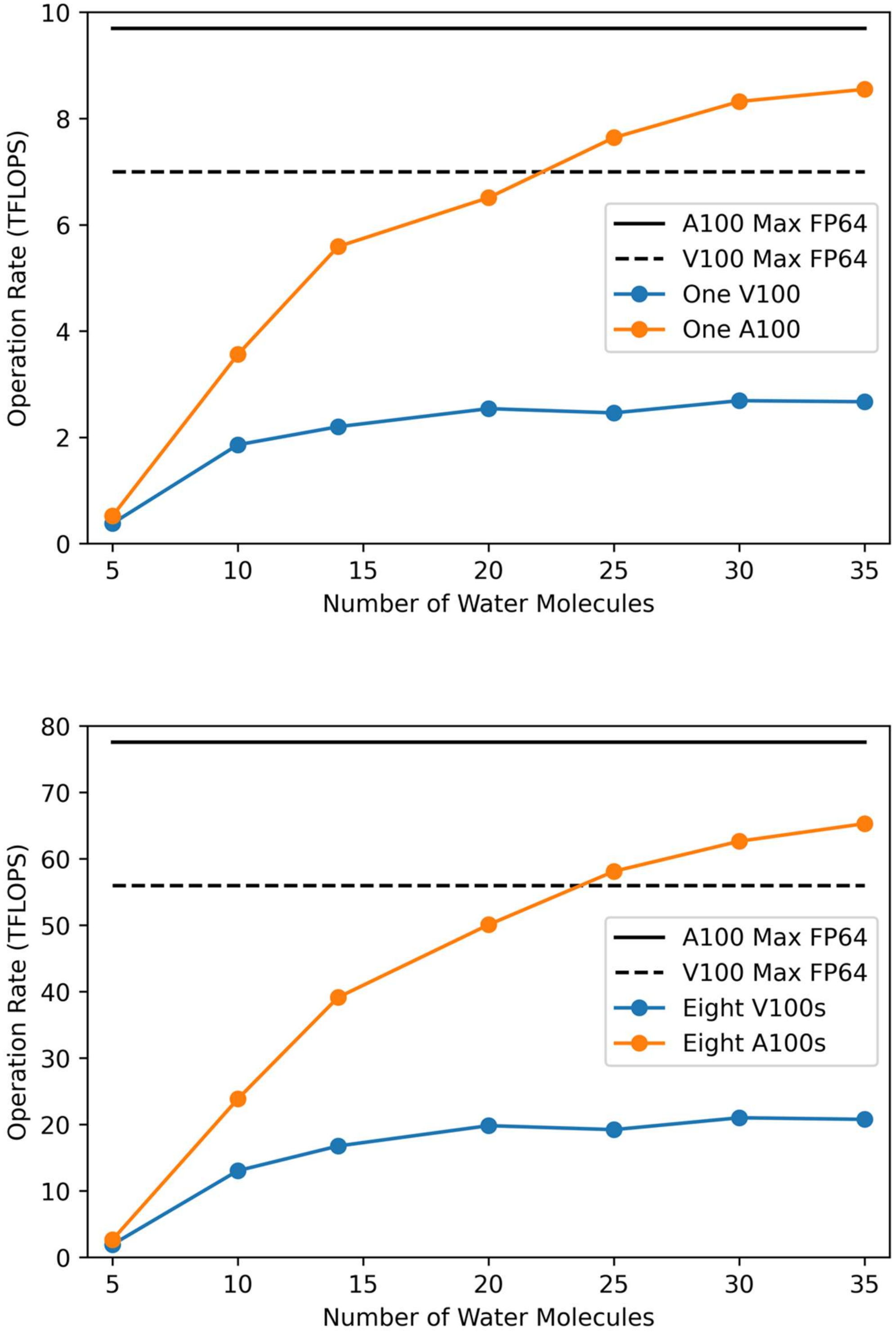

**Figure S1.** Operation rate as measured in TFLOPS as a function of system size (number of water molecules with a cc-pVDZ basis set[6, 7]). The theoretical peak performance of 1 A100 is shown by the solid black line, while that of 1 V100 is shown by the dashed line (top). The same plot for 8 A100s/V100s is shown in the bottom panel.

References

1. Peng, C.; Calvin, J. A.; Valeev, E. F., Coupled-cluster singles, doubles and perturbative triples with density fitting approximation for massively parallel heterogeneous platforms. *Int. J. Quantum Chem.* **2019,** *119*, e25894.
2. Gyevi-Nagy, L.; Kállay, M.; Nagy, P. R., Integral-Direct and Parallel Implementation of the CCSD(T) Method: Algorithmic Developments and Large-Scale Applications. *J. Chem. Theory Comput.* **2020,** *16*, 366.
3. Datta, D.; Gordon, M. S., A Massively Parallel Implementation of the CCSD(T) Method Using the Resolution-of-the-Identity Approximation and a Hybrid Distributed/Shared Memory Parallelization Model. *J. Chem. Theory Comput.* **2021,** *17*, 4799.
4. Shen, T.; Zhu, Z.; Zhang, I. Y.; Scheffler, M., Massive-Parallel Implementation of the Resolution-of-Identity Coupled-Cluster Approaches in the Numeric Atom-Centered Orbital Framework for Molecular Systems. *J. Chem. Theory Comput.* **2019,** *15*, 4721.
5. Anisimov, V. M.; Bauer, G. H.; Chadalavada, K.; Olson, R. M.; Glenski, J. W.; Kramer, W. T. C.; Aprà, E.; Kowalski, K., Optimization of the Coupled Cluster Implementation in NWChem on Petascale Parallel Architectures. *J. Chem. Theory Comput.* **2014,** *10*, 4307.
6. Kendall, R. A.; Dunning, T. H.; Harrison, R. J., Electron Affinities of the First-Row Atoms Revisited. Systematic Basis Sets and Wave Functions. *J. Chem. Phys.* **1992,** *96*, 6796.
7. Dunning, T. H., Gaussian Basis Sets for Use in Correlated Molecular Calculations. I. The Atoms Boron through Neon and Hydrogen. *J. Chem. Phys.* **1989,** *90*, 1007.